\documentclass{moriond}
\usepackage{bm}
\def\Journal#1#2#3#4{{#1} {\bf #2}, #3 (#4)}

\def\ApJ{\em Astrophysical Journal}

\def\CQG{\em Classical and Quantum Gravity}
\def\NewA{\em New Astronomy}

\def\PRD{{\em Phys. Rev.} D}

\def\be{\begin{equation}}
\def\ee{\end{equation}}
\def\bea{\begin{eqnarray}}
\def\eea{\end{eqnarray}}
\def\bi{\begin{itemize}}
\def\ei{\end{itemize}}
\newcommand{\Mbh}{M_\bullet}
\newcommand{\Mo}{M_\odot}
\newcommand{\Ms}{M_\star}
\newcommand{\Ns}{N_\star}

\newcommand{\Photo}{}

\begin{document}
\vspace*{4cm}
\title{RELATIVISTIC STELLAR DYNAMICS AROUND A MASSIVE BLACK HOLE IN STEADY STATE}

\author{TAL ALEXANDER}

\address{Department of Particle Physics and Astrophysics, Weizmann
  Institute of Science, Rehovot 76100, Israel}

\maketitle

\abstracts{I briefly review advances in the understanding and modeling
  of relativistic stellar dynamics around massive black holes (MBHs)
  in galactic nuclei, following the inclusion of coherent relaxation
  and of secular processes in a new formal analytic description of the
  dynamics.}

\section{Relaxation in galactic nuclei}
\label{s:intro}
The dense, centrally concentrated stellar cluster that exists around
most MBHs offers opportunities for strong, possibly destructive
interactions between it and stars. These include direct plunges,
leading to tidal disruption flares or gravitational waves (GW) flares,
inspiral processes leading to quasi-periodic GW emission from extreme
mass ratio inspiral events (EMRIs)~\cite{ama+07}, tidally powered
stars (``squeezars'')~\cite{ale+03a}, strong tidal
scattering~\cite{ale+01} or capture by massive accretion disks. These
processes affect MBH growth and may create exotic stellar populations
around MBHs~\cite{ale+01}.

This naturally leads to the question ``How do stars closely interact
with, and fall into a MBH, and at what rates?'' This is known as the
stellar dynamical ``loss-cone problem''. It is a non-trivial problem,
in spite of the presence of so many stars so close to the MBH, because
the phase space volume of unstable orbits is minute. The few stars
initially on such orbits quickly fall into the MBH on the short
dynamical timescale, and then the rates would drop to zero, if it were
not for dynamical processes that deflect additional stars from stable
orbits to those with velocity vectors that point toward the MBH,
within the loss-cone (Fig. \ref{f:LC} left). Thus, the loss-cone
question is essentially the question: ``how do galactic nuclei
randomize and relax?''

\subsection{Non-coherent 2-body relaxation (NR)}
\label{ss:NR}
\begin{figure}
\begin{minipage}{0.3\columnwidth}
\centerline{\includegraphics[width=1.1\linewidth]{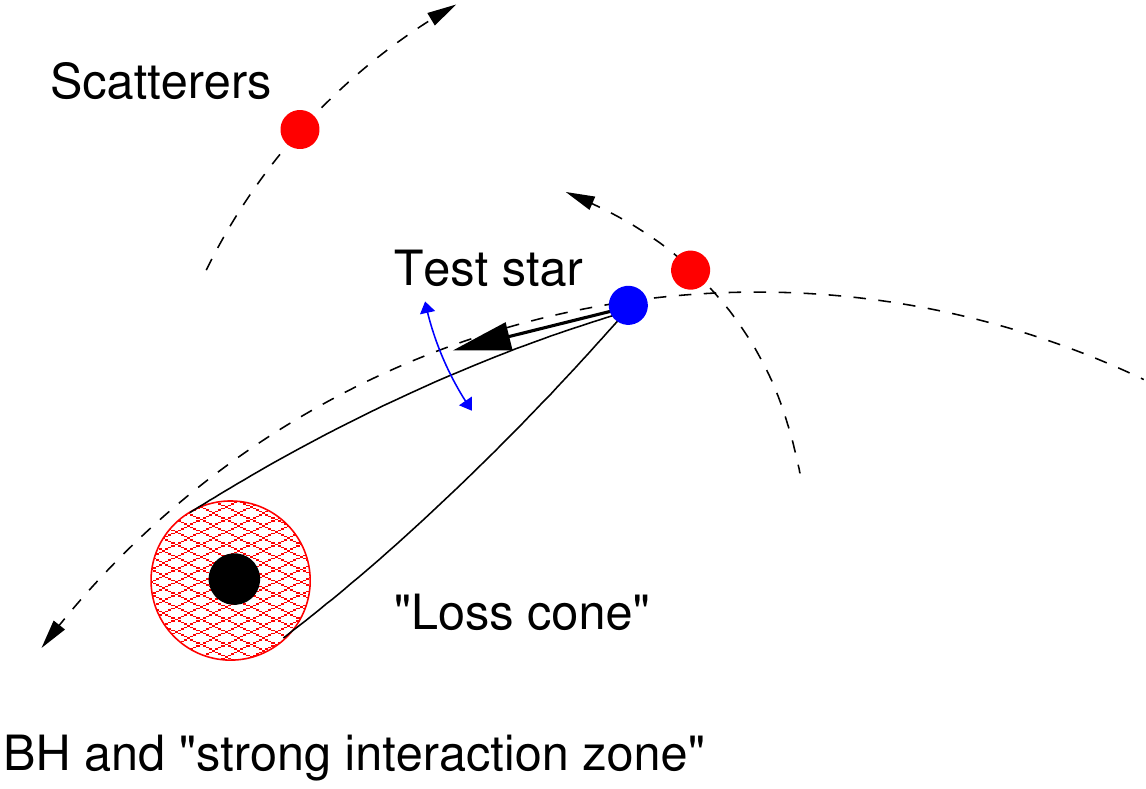}}
\end{minipage}
\hfill
\begin{minipage}{0.3\linewidth}
\centerline{\includegraphics[width=1.1\linewidth]{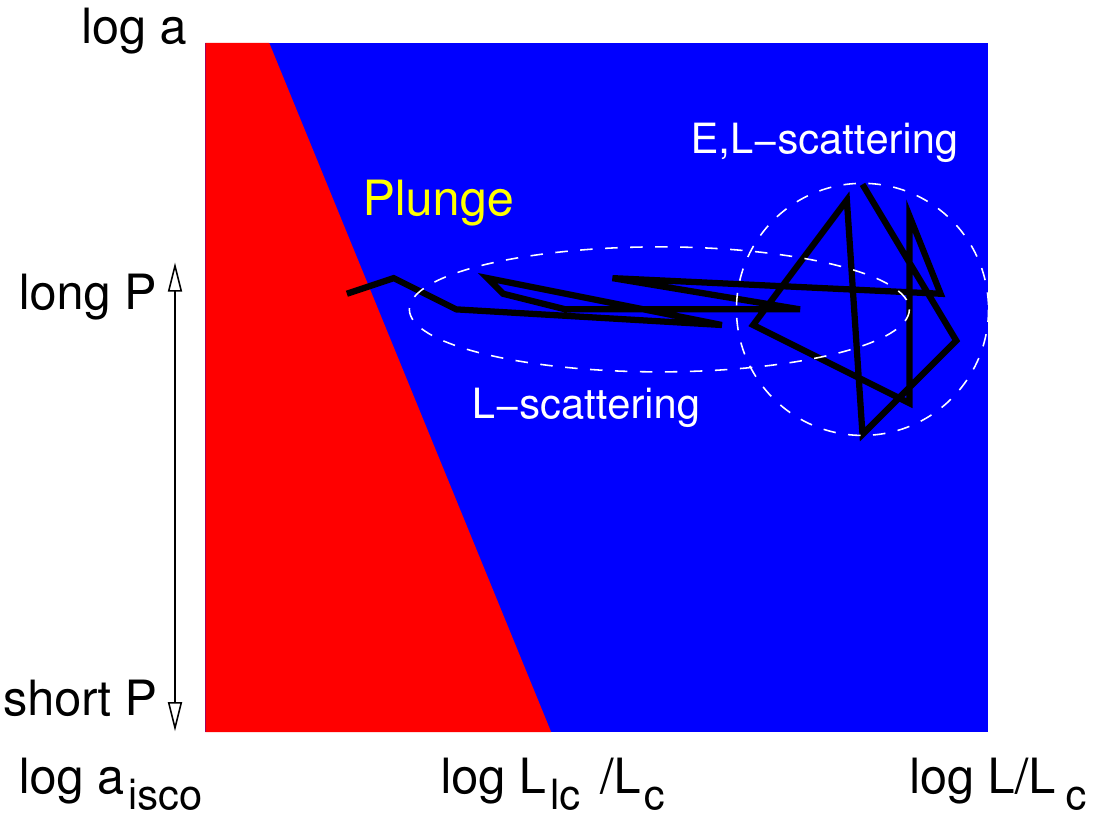}}
\end{minipage}
\hfill
\begin{minipage}{0.3\linewidth}
\centerline{\includegraphics[width=1.1\linewidth]{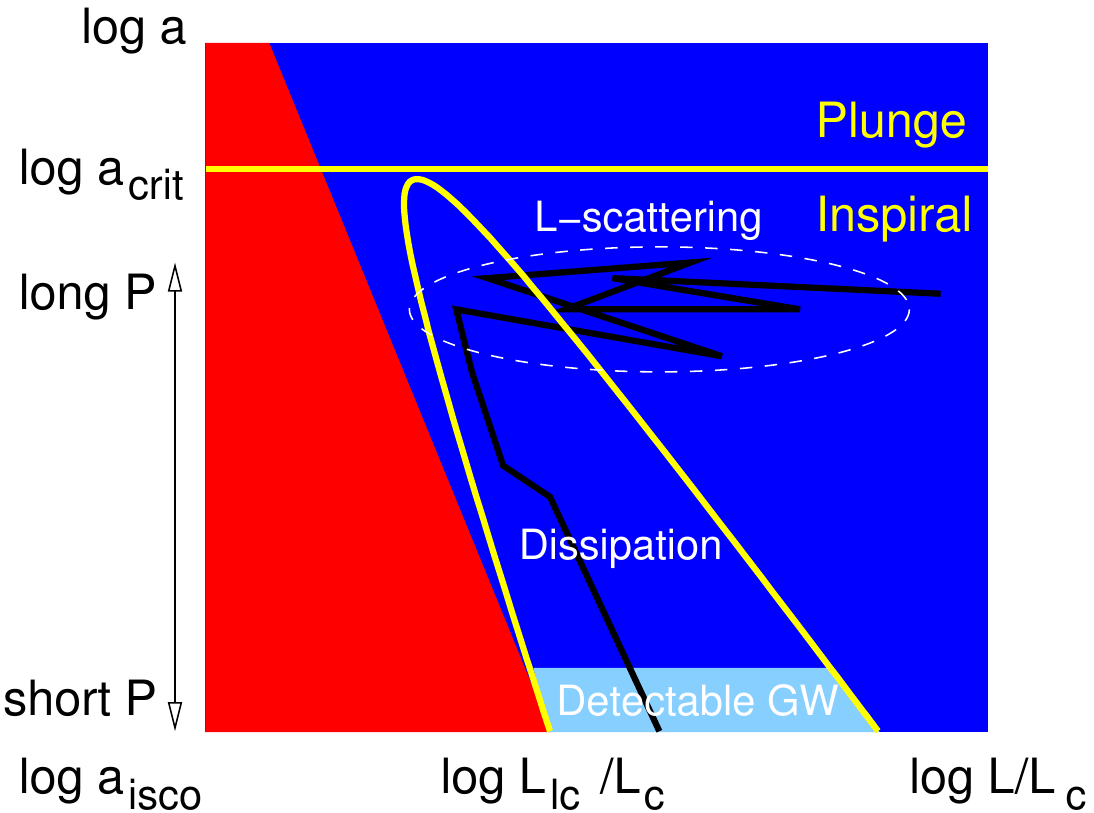}}
\end{minipage}
\caption[]{\label{f:LC}
  The 2-body relaxation-driven loss-cone. {\bf Left}: A star
  is scattered to an orbit in the loss-cone, which takes it close
  enough to the MBH for a strong (possibly destructive) interaction
  with it. {\bf Center}: The loss-cone phase space in terms of the
  normalized angular momentum $j=L/L_c=\sqrt{1-e^2}$ and the
  semi-major axis $a$, without dissipation. {\bf Right}: The same, but
  with a dissipative mechanism (here, the emission of GWs). See text.}
\end{figure}

The discreteness of stellar systems leads to non-coherent 2-body
relaxation (NR) (Fig. \ref{f:LC} left). This guarantees a minimal
relaxation rate, on a timescale $T_{NR} \sim
Q^2P(r)/\Ns(r)\log Q$, where $Q=\Mbh/\Ms$ is the MBH/star mass ratio,
$P(r)$ is the radial orbital period, and $\Ns(r)$ the number of stars
inside $r$. Because the impact parameter $b$ of these {\em point-point
  interactions} can be small, NR is boosted by the Coulomb factor
$\log(b_{\max} / b_{\min})=\log Q$. $T_{NR}$ is the timescale for
changes of order unity in energy, $T_E$. It is however easier to drive
a star into the MBH by reducing its angular momentum $L$ and making
its orbit more radial, than by reducing the orbital energy $E<0$, and
shrinking the orbit. The timescale for changing
$j=L/L_c(a)=\sqrt{1-e^2}$ from $j$ to 0 is $T_L=j^2T_E$
($L_c=\sqrt{G\Mbh a}$ is the circular $L$, $a$ the sma and $e$ the
eccentricity).

In the absence of dissipation, stars with $j\ll 1$ are deflected by
$L$-scattering at nearly constant $a$ to the innermost stable orbit
(ISO), at $j_{\rm iso}=4\sqrt{r_g/a}$ ($r_g=G\Mbh/c^2$) and then plunge
directly into the MBH~\cite{lig+77} (Fig. \ref{f:LC} center). When a
dissipative mechanism is present (e.g. GW), phase space is divided in
two (Fig. \ref{f:LC} left). Below some critical sma $a_c$, all stars
eventually cross the ``inspiral line'' where $E$-dissipation is faster
than $j$-scattering, and then inspiral gradually into the MBH as
EMRIs. Stars above $a_c$ plunge directly. The respective rates of
plunges and inspirals can then by estimated by the ratio of number of
stars on the relevant scales (the MBH radius of influence $r_h$ for
plunges (e.g tidal disruptions), and $a_c$ for inspirals) over
$T_{NR}$ on that scale: $R_p\sim \Ns(r_h)/T_{NR}\log(L_c/L_{\rm iso})$
and $R_i\sim \Ns(a_c)/T_{NR}\log(L_c/L_{\rm iso})$. Because
$\Ns(a_c)\ll \Ns(r_h)$, the inspiral rate is  much lower than
the plunge rate~\cite{ale+03b}, typically $R_i\sim 0.01R_p$.

\subsection{Coherent resonant relaxation (RR) in nearly-spherical systems}
\label{ss:RR}
Resonant relaxation~\cite{rau+96} is a process of rapid $L$-relaxation
that occurs when the gravitational potential is symmetric enough to
restrict the evolution of orbits on timescales much longer than the
orbital time (e.g. nearly-fixed Keplerian ellipses in the
nearly-Keplerian potential close to a MBH, where the stellar mass is
negligible, but far enough so that GR effects are weak). In that case,
a test orbit will feel a residual torque from the static,
orbit-averaged background of stellar ``mass wires'', which persists
for a long coherence time $T_c$, until small deviations from symmetry
accumulate and randomize the background. These {\em orbit-orbit
  interactions} randomize the angular momentum on a timescale $T_{RR}
\sim Q^2P(r)^2/\Ns(r)T_c$. Unlike NR, these extended objects do not
undergo close interactions. Rather, RR is boosted by the long
coherence time.

\begin{figure}
\begin{minipage}{0.3\columnwidth}
\centerline{\includegraphics[width=0.9\linewidth]{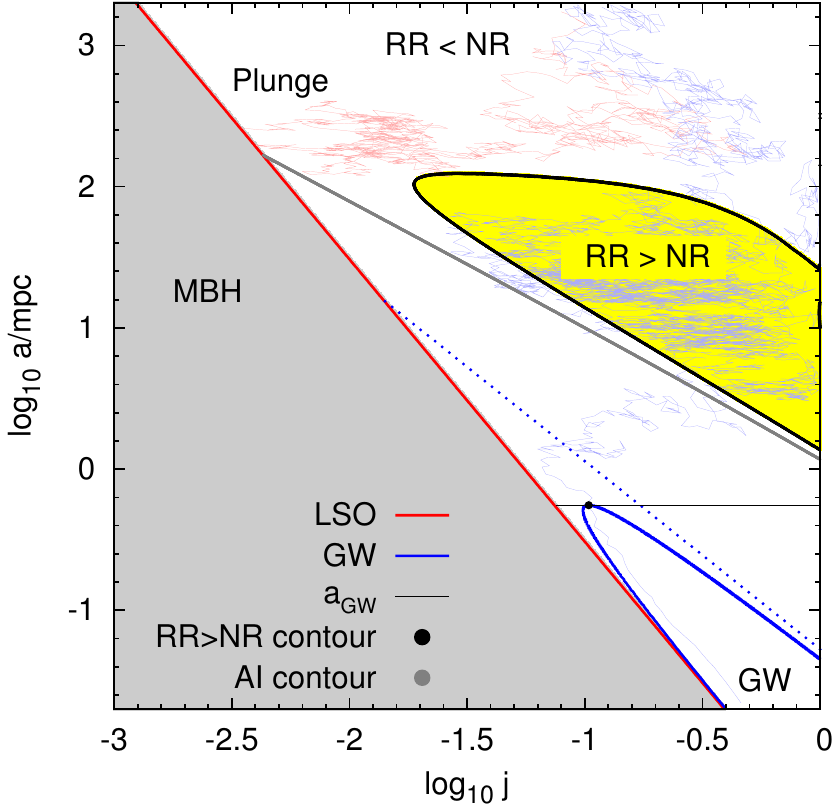}}
\end{minipage}
\hfill
\begin{minipage}{0.3\linewidth}
\centerline{\includegraphics[width=1.14\linewidth]{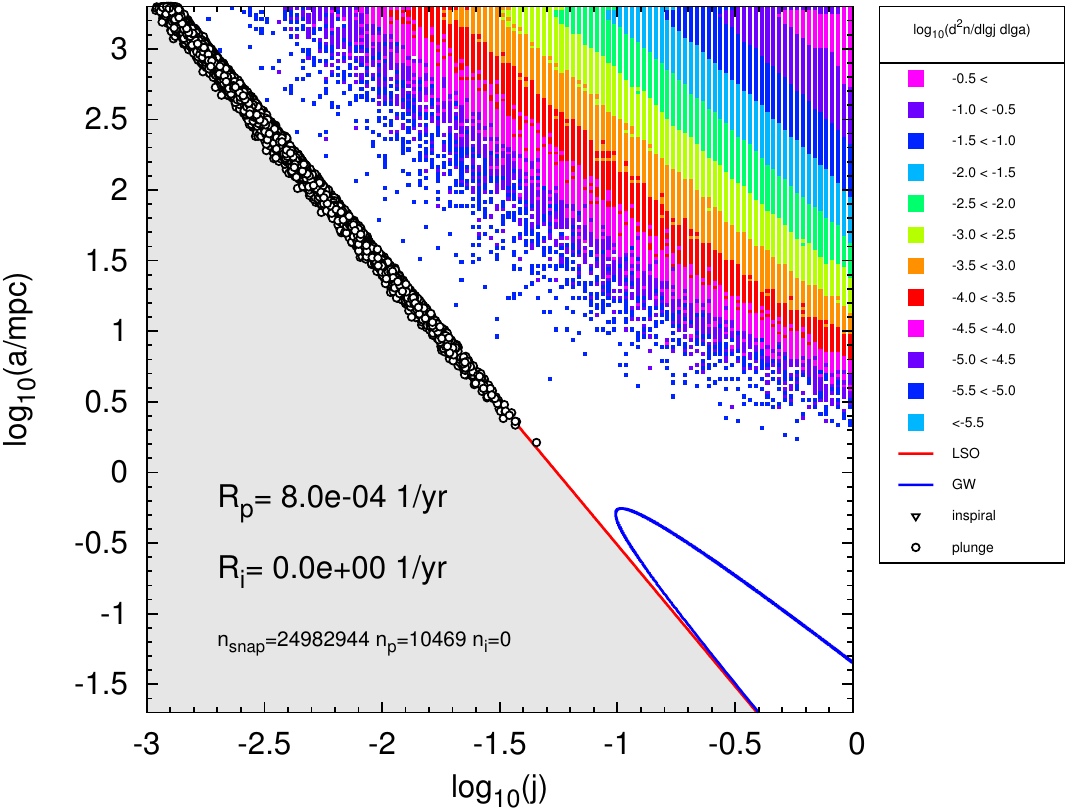}}
\end{minipage}
\hfill
\begin{minipage}{0.3\linewidth}
\centerline{\includegraphics[width=1.14\linewidth]{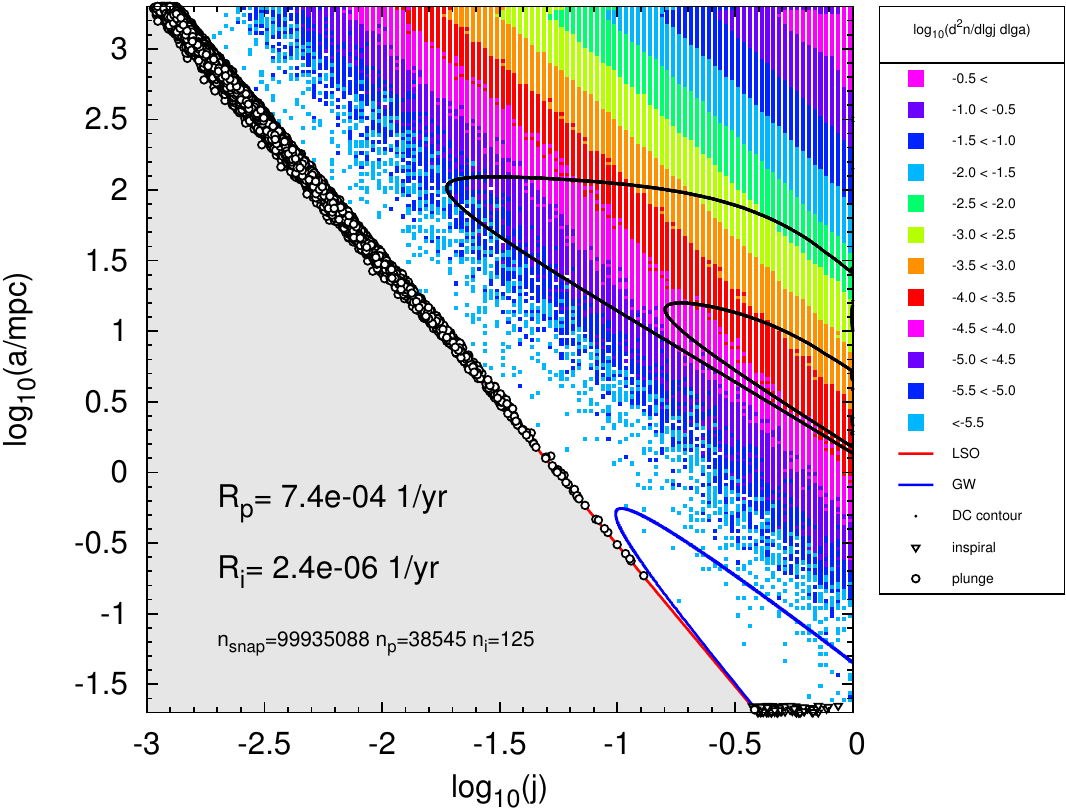}}
\end{minipage}
\caption[]{\label{f:EJ} The loss-cone phase space. {\bf Left}: A
  schematic for a model of the Milky Way nucleus
  ($\Mbh=4\times10^6\,\Mo$, $\Ms=10\,\Mo$). Orbits are unstable in the
  gray region left of the ISO line. Dynamics are dominated by GW
  dissipation inside the GW curve at the bottom right. The horizontal
  line tangent to the upper tip of the GW line is the critical sma
  separating plunge and inspiral tracks (one example shown for
  each). RR dominates over NR in the shaded region in the middle
  right. RR is ineffective on short timescales below the diagonal AI
  line just below the RR region. {\bf Center}: The phase space density
  and loss rates calculated by MC simulations, for an artificial model
  without GR precession to quench RR, resulting in rapid plunges
  (circles at ISO line) and complete suppression of EMRIs. {\bf
    Right}: The same, but with GR precession, which suppresses RR (RR
  remains strong only inside the $T_{RR}=0.1,1 T_{NR}$ contours). This
  enables EMRIs (triangles at bottom of GW region). See text.}
\end{figure}

RR is relevant for the loss-cone problem because it is possible to
have $T_c\gg P$ in the symmetric potential near a MBH, so that
$T_{RR}/T_{NR}\sim (\log Q)P/T_c\ll 1$. That is, angular momentum
evolution, and in particular that leading to $j\to 0$ and strong
interactions with the MBH, can be greatly accelerated. Unchecked, RR
will completely suppress EMRIs by driving all stars into plunge orbits
(Fig. \ref{f:EJ} center). However, very eccentric orbits undergo GR
in-plane (Schwarzschild) precession, which quenches RR by rapidly
alternating the direction of the residual torque on the orbit. This
motivated the ``fortunate coincidence conjecture''~\cite{hop+06}: The
${\cal O}(\beta^2j^{-2})$ GR precession becomes significant before
${\cal O}(\beta^5j^{-7}Q^{-1})$ GW dissipation, and this may allow EMRIs
to proceed unperturbed, decoupled from the background stars.

\subsection{The Schwarzschild Barrier}
\label{ss:SB}
The first full PN2.5 $N$-body simulations~\cite{mer+11} revealed a
surprising result: not only does GR precession quench RR before the
GW-dominated regime, as conjectured, but there appears to be some kind
of barrier in phase space, dubbed the Schwarzschild Barrier (SB),
which prevents the orbits from evolving to $j\to0$. Instead, they
appear to linger for roughly $T_c$ near the SB, where their orbital
parameters oscillate at the GR precession frequency, and then they
evolve back to $j\to1$. An early analysis~\cite{ale10} suggested that
this behavior is related to precession under the influence of a
residual dipole-like residual force. However, a full self-consistent
explanation of the SB was lacking, and its very existence and nature
remained controversial.

I now describe briefly a new formal framework for expressing coherent
relaxation and secular processes in galactic nuclei~\cite{bar+14}, and
discuss implications for the steady state phase space structure and
loss rates~\cite{bar+15}.

\section{Hamiltonian dynamics with correlated background noise}
\label{s:eta}

Two key insights inform the new advances in understanding coherent
relativistic dynamics, which lie in the difficult-to-treat interface
between deterministic Hamiltonian dynamics and stochastic kinetic
theory. (1) The effect of the background stars on a test stars should
be described by a {\em correlated} noise model $\bm{\eta}(t)$, whose
degree of smoothness (differentiability) determines dynamics on short
timescales~\cite{bar+14}. (2) The long-term steady-state remains
(unavoidably) the maximal entropy configuration, irrespective of the
details of the nature of the relaxation processes~\cite{bar+15}.

This formal treatment of PN1 dynamics in the presence of correlated
(RR) noise $\bm\eta$ (a 3-vector in $L$-space) allows to write a
phase-averaged leading-order ($\ell=1$) Hamiltonian $\bar{{\cal H}}_1$
and derive {\em stochastic} EOMs for the orbital elements of a test
star, $\bm{x}\equiv(j,\phi,\cos\theta)$ and the argument of periapse
$\psi$, which precesses at frequency $\nu_p(j)$, \be \dot{\bm{x}} =
\bm{\nu}_{\tau,\bm{x}}(\bm{x},\psi)\cdot\bm{\eta}\,,\,\,\,\,\,\,\,\,\,
\dot{\psi} =
\bm{\nu}_{\tau,\psi}(\bm{x},\psi)\cdot\bm{\eta}+\nu_p(j)\,,
\ee
where $\nu_\tau$ is the RR torque frequency. This $\bm\eta$ formalism
allows to evolve a test star in time for a given realization of the
noise. Moreover, even though $\bm\eta$ is time-correlated, it is possible
to derive (and validate with the stochastic EOMs) approximate
diffusion coefficients (DCs) $D_{1,2}$, which allow to evolve in time
the probability density $\rho(j)$ with the Fokker-Planck (FP)
equation,
\be
\frac{\partial\rho}{\partial t} = \frac{1}{2}\frac{\partial}{\partial j}\left\{
jD_2\frac{\partial}{\partial j}\left[\frac{\rho}{j} \right]
\right\}\,, \,\,\,{\rm where}\,\,\,\,
D_2=\left|\bm{\nu}_{\tau,j}\right|^2{\cal F}_{C(t)}[\nu_p(j)]
\,\,\,\,{\rm and}\,\,\,\, D_1=\frac{1}{2j}\frac{\partial
  jD_2}{\partial j}\,.
\ee
${\cal F}_{C(t)}$ is the Fourier transform of $\bm\eta$'s
auto-correlation function (ACF). The explicit dependence of $D_2$ on
the spectral power of the noise at the precession frequency is an
expression of {\em adiabatic invariance} (AI). If, {\em and only if}
the noise has an upper frequency cutoff, as it must if it is
smooth (this is physically expected, since the background noise
is the superposition of continuous orbital motions), then there is a
critical $j_0$ such that for $j<j_0$ the precession is fast enough so
that $D_2(j)\to 0$, and the star decouples from the background
resonant torques (Fig. \ref{f:eta} left, center). This describes the
dynamics of the SB: it is not a reflecting boundary, but a locus in
phase space where diffusion rapidly drops due to AI. Since diffusion
to yet lower $j$ slows further down, while diffusion to higher $j$
speeds further up, orbits statistically seem to bounce away from the SB.

\section{The steady state loss-cone}
\label{s:SSLC}
\begin{figure}
\begin{minipage}{0.3\columnwidth}
\centerline{\includegraphics[width=1.1\linewidth]{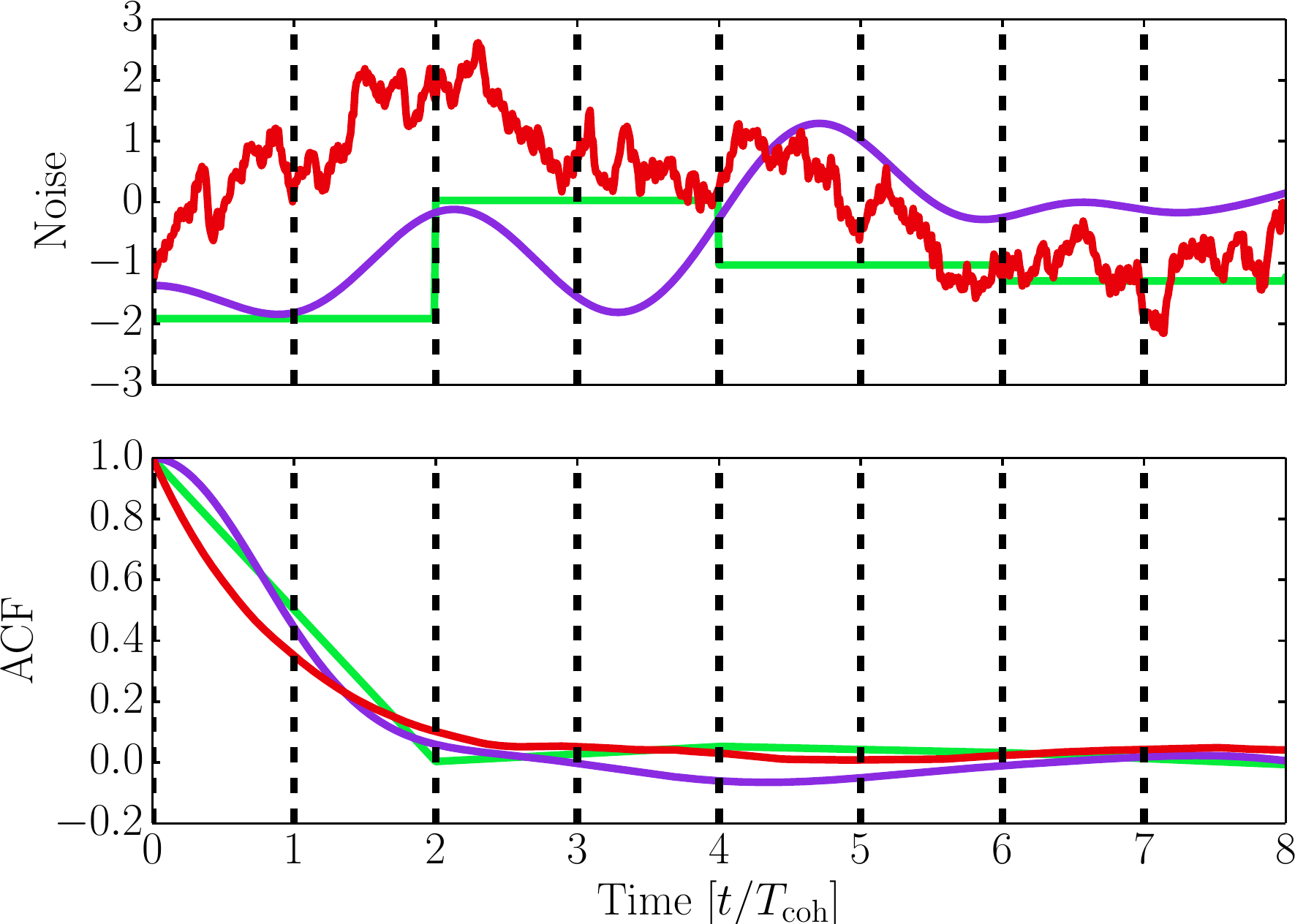}}
\end{minipage}
\hfill
\begin{minipage}{0.3\linewidth}
  \centerline{\includegraphics[width=1.1\linewidth]{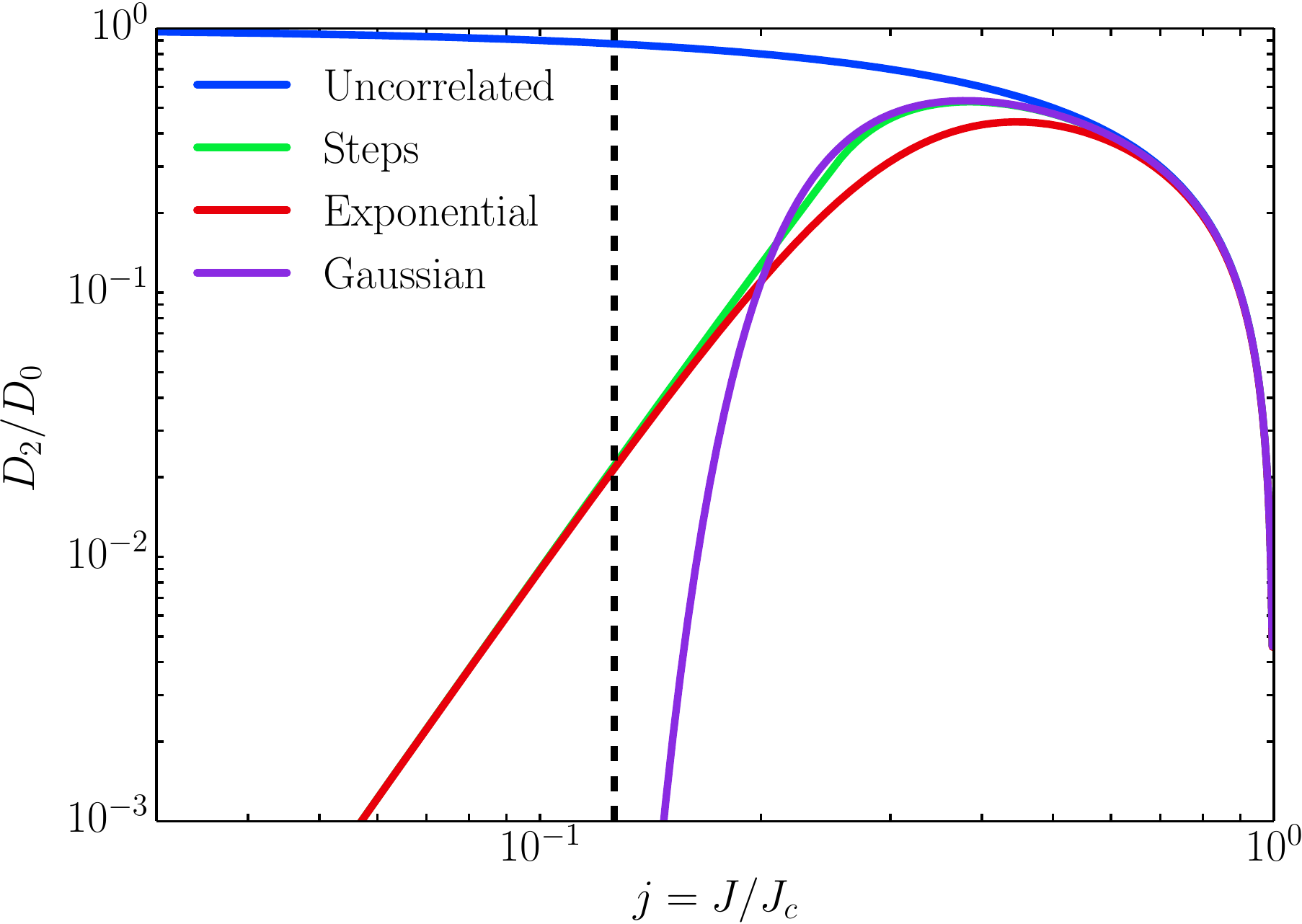}}
\end{minipage}
\hfill
\begin{minipage}{0.3\linewidth}
\centerline{\includegraphics[width=1.1\linewidth]{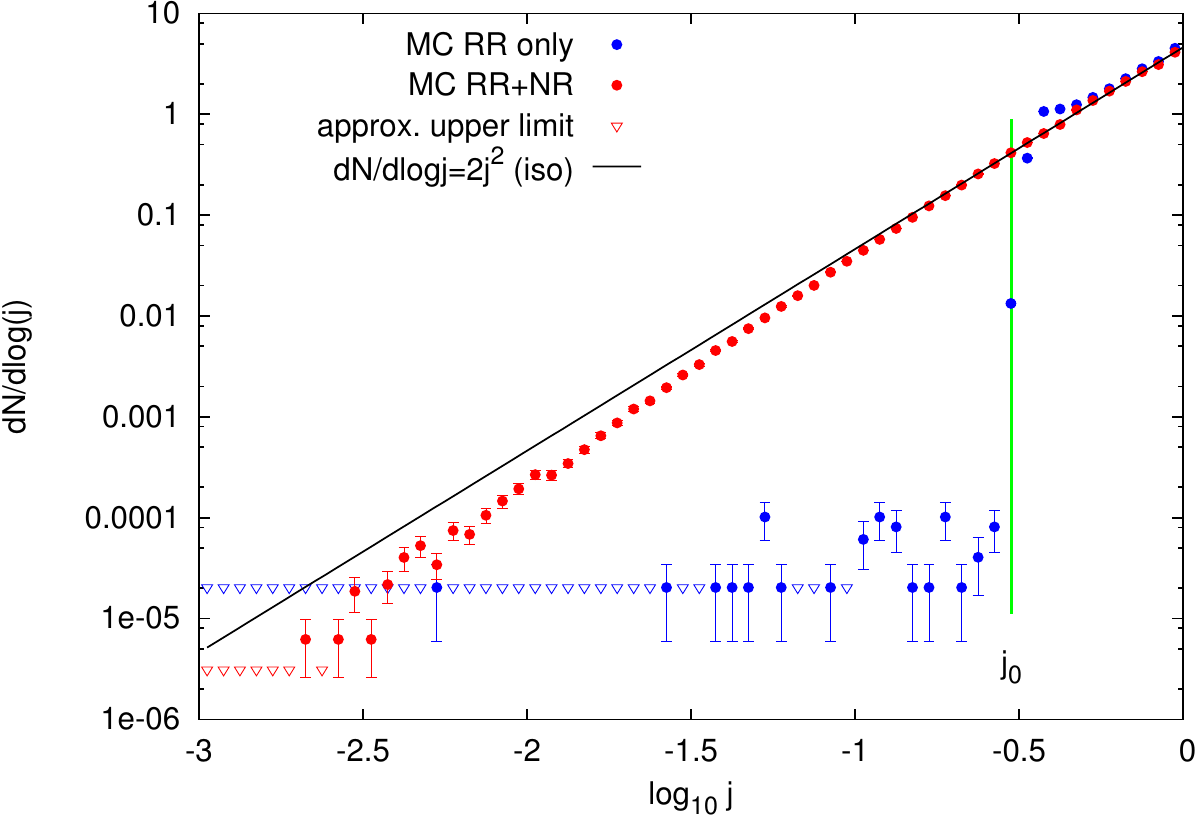}}
\end{minipage}
\caption[]{\label{f:eta} The smoothness of the noise model and
  diffusion dynamics.  {\bf Left}: Three $\bm\eta$ models and their
  ACF: discontinuous steps ($C^0$,), continuous but not continuously
  differentiable ($C^1$ with exponential ACF), smooth ($C^\infty$ with
  Gaussian ACF). {\bf Center}: The corresponding $D_2$; note the steep
  cutoff at $j<0.1$ for the smooth noise model. {\bf Right}: The MC
  simulations of $j$-only evolution reproduce the AI/SB limit at
  $j<j_0$ in the absence of NR, but NR erases this feature completely
  on timescale $t\to T_{NR}$.}
\end{figure}

NR is impervious to AI. When $t\to T_{NR}$, the system approaches the
maximal entropy solution ($dN/dj=2j$, Fig. \ref{f:eta} right). Monte
Carlo (MC) simulations of the probability density, branching ratios
and loss rates with the $\bm\eta$ formalism (Fig. \ref{f:EJ} right) show
that the RR-dominated region in phase space is well separated from the
plunge and inspiral loss-lines, so the effect of RR on the loss rates
is small ($< \times2-3$). Specifically, we conclude that GR quenching
of RR is effective, so the EMRI rates remain largely unaffected by RR. RR
can be significant for processes whose loss-line crosses the
RR-dominated region, e.g. destruction by interaction with an accretion
disk.

\section{summary}
\label{s:summary}
NR, RR, GW dissipation and secular precession can be treated
analytically as effective diffusion with correlated noise. The steady
state depends mostly on NR, which erases AI. RR can be important in
special cases. The $\bm\eta$ formalism provides stochastic EOMs for
evolving test particles and an FP/MC diffusion procedure for evolving
the probability density. This makes it possible to model the
relativistic loss-cone in galactic nuclei with realistic $\Ns\gg1$,
(unlike direct $N$-body simulations), and obtain the branching ratios,
loss rates and steady state stellar distributions.

\section*{Acknowledgments}

This research was supported by the I-CORE program of the PBC and ISF
(grant No 1829/12).

\end{document}